\documentclass[
    ,final            % use final for the camera ready runs
  ]
  {aipproc}

\layoutstyle{8x11double}

\makeatletter

\def\vereq#1#2{\lower3pt\vbox{\baselineskip1.5pt \lineskip1.5pt
\ialign{$\m@th#1\hfill##\hfil$\crcr#2\crcr\sim\crcr}}}
\def\lesssim{\mathrel{\mathpalette\vereq<}}
\@addtoreset{equation}{section} \makeatother

\begin{document}

\title{New bounce solutions and vacuum tunneling in de Sitter 
spacetime}


\author{Erick J. Weinberg}{
  address={Department of Physics, Columbia University, New York, NY
  10027, USA}
}

\begin{abstract}
I describe a class of oscillating bounce solutions to the Euclidean
field equations for gravity coupled to a scalar field theory with
multiple vacua.   I discuss their implications for vacuum
tunneling transitions and for elucidating the thermal nature of 
de Sitter spacetime.

\end{abstract}

\maketitle

\section{Introduction}

The problem of vacuum tunneling in de Sitter spacetime has recently
acquired renewed relevance.  In part, this is due to developments in
string theory, which suggest that vacuum tunneling may be of relevance
for understanding transitions between the various potential vacua that
populate the string theory landscape.  But, it is also of interest for
the light that it can shed on the nature of de Sitter spacetime.  In
this talk I will describe some recent work~\cite{Hackworth} with Jim
Hackworth in which we explored some aspects of the
subject that have received relatively little attention.

De Sitter spacetime is the solution to Einstein's equations when there
is a constant positive vacuum energy density $V_{\rm vac}$, but no other
source.  Globally, it can be represented as the hyperboloid $x^2 + y^2
+ z^2 + w^2 -v^2 = H^{-2}$ in a flat five-dimensional space with
metric $ds^2 = dx^2 + dy^2 + dz^2 + dw^2 -dv^2$,  where
\begin{equation}
   H^2 = {8 \pi \over 3} {V_{\rm vac}  \over M_{\rm Pl}^2 }  \, .
\end{equation}
The surfaces of constant $v$ are three-spheres, with the sphere of
minimum radius, $H^{-1}$, occurring at $v=0$.  However, the special
role played by this surface is illusory.  De Sitter spacetime is
homogeneous, and a spacelike three-sphere of minimum radius can be
drawn through any point.

An important property of de Sitter spacetime is the existence of horizons.
Just as for the case of a black hole, the existence of a
horizon gives rise to thermal radiation, characterized by a temperature
\begin{equation}
    T_{\rm dS} = H/2\pi \, .
\end{equation}
However, there are important differences from the black hole case.
A black hole horizon has a definite location, independent of
the observer.  Further, although an observer's motion affects how the
thermal radiation is perceived, the radiation has an unambiguous,
observer-independent consequence --- after a finite time, the black
hole evaporates.  By contrast, the location of the de Sitter horizon
varies from observer to observer.  Although comoving observers
detect thermal radiation with a temperature $T_{\rm dS}$, this
radiation does not in any sense cause the de Sitter spacetime to
evaporate.  As we will see, tunneling between different de Sitter
vacua provides further insight into the thermal nature of de Sitter
spacetime and the meaning of $T_{\rm dS}$.

Of course, the relevance of de Sitter spacetime to our Universe comes
from the fact that in the far past, during the inflationary era, and
in the far future, if the dark energy truly corresponds to a
cosmological constant, the Universe approximates a portion of de
Sitter spacetime.  The underlying assumption is that results derived
in the context of the full de Sitter spacetime are applicable to 
a region that is approximately de Sitter over a spacetime volume
large compared to $H^{-4}$.

\begin{figure}[b]
  \includegraphics[height=.22\textheight]{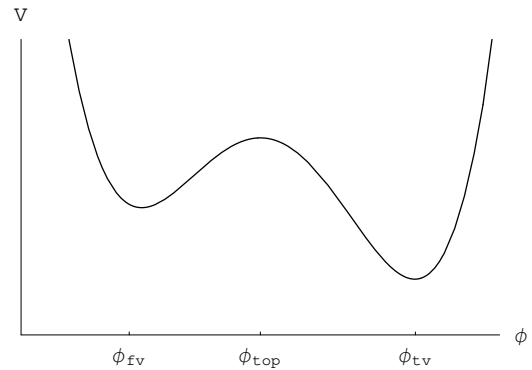}
  \caption{The potential for a typical theory with a false vacuum}
\label{potentialfig}
\end{figure}

\section{Vacuum decay in flat spacetime}

The prototypical example for studying vacuum decay is a scalar field
theory with a potential, such as that shown in
Fig.~\ref{potentialfig}, that has both an absolute minimum (the ``true
vacuum'') and a higher local minimum (the ``false vacuum'').  For the
purposes of this talk I will assume that $V(\phi) > 0$, so that both
vacua correspond to de Sitter spacetimes.  The false vacuum is a
metastable state that decays by quantum mechanical tunneling.  
It
must be kept in mind, however, that the tunneling is not
from a homogeneous false vacuum to a homogeneous true vacuum, as
might be suggested  by the plot of $V(\phi)$.  Rather,
the decay proceeds via bubble nucleation, with tunneling being from
the homogeneous false vacuum to a configuration containing a bubble of
(approximate) true vacuum embedded in a false vacuum background.
After nucleation, the bubble expands, a classically allowed
process.

I will begin the discussion of this process by recalling the simplest
case of quantum tunneling, that of 
a point particle of mass $m$ tunneling through a one-dimensional
potential energy barrier $U(q)$ from a initial point $q_{\rm init}$ to
a point $q_{\rm fin}$ on the other side of the barrier.
The WKB approximation 
gives a tunneling rate proportional to $e^{-B}$,
where
\begin{equation}
      B = 2 \int_{q_{\rm init}}^{q_{\rm fin}} 
               dq \sqrt{2m[U(q)- E]} \, .
\label{WKBfactor}
\end{equation}
This result can be generalized to the case of 
a multi-dimensional system with coordinates $q_1, q_2, \dots, q_N$.
Given an initial point $q_j^{\rm init}$, one considers paths
$q_j(s)$ that start at $q_j^{\rm init}$ and end at some point
$q_j^{\rm fin}$ on the opposite side of the barrier.  
Each such path defines a one-dimensional tunneling integral $B$.
The WKB tunneling exponent is obtained from the path that
minimizes this integral~\cite{Banks:1973ps}
As a bonus, this minimization process also 
determines the optimal exit point from the barrier. 

By manipulations analogous to those used in classical mechanics (but
with some signs changed), this minimization problem
can be recast as the problem of finding a stationary point of the
Euclidean action
\begin{equation}
   S_{\rm E} = \int_{\tau_{\rm init}}^{\tau_{\rm fin}} 
    d\tau \left[
      {m\over 2} \left({dq^j\over d\tau}\right)^2 + U(q)\right]\, .
\end{equation}
One is thus led to solve the Euclidean equations of motion
\begin{equation}
    0 = m {d^2q_j \over d\tau^2} + {\partial U \over \partial q_j}\, .
\end{equation}
The boundary conditions are that $q_j(\tau_{\rm init}) = q_j^{\rm
init}$ and (because the kinetic energy vanishes at the point where the 
particle emerges from the barrier)
that $dq_j/d\tau = 0$ at $\tau_{\rm fin}$.  
The
vanishing of $dq_j/d\tau$ at the endpoint implies that the solution can be
extended back, in a ``$\tau$-reversed'' fashion, to give a solution
that runs from $q_j^{\rm init}$ to $q_j^{\rm fin}$ and back again
to $q_j^{\rm init}$.  This solution is known as a ``bounce'', and 
the tunneling exponent is given by 
\begin{eqnarray}
   B  \!\! \!\! \!\!&=& \!\! \!\! \!\! \int    d\tau \left[
      {m\over 2} \left({dq^j\over d\tau}\right)^2 + U(q)
         -U(q^{\rm init})  \right] \cr
     \!\! \!\! \!\!&=& \!\! \!\! \!\! S_{\rm E}({\rm bounce}) 
          - S_{\rm E}({\rm false~vacuum})  \, ,
\end{eqnarray}
with the factor of 2 in Eq.~(\ref{WKBfactor}) being absorbed by the
doubling of the path.  It is essential to remember that
$\tau$ is not in any sense a time, but merely one of many possible
parameterizations of the optimal tunneling path.

The translation of this to field theory~\cite{ColemanI} is
straightforward: The coordinates $q_j$ become the field variables
$\phi({\bf x})$, and the path $q_j(\tau)$ becomes a series of
three-dimensional field configurations $\phi({\bf x}, \tau)$.  The
Euclidean action is
\begin{equation}
    S_{\rm E} = 
  \int d\tau \, d^3{\bf x} \left[ 
     {1\over 2} \left({\partial \phi\over \partial \tau}\right)^2
     +{1\over 2} ({\bf \nabla}\phi)^2 + V(\phi) \right]
\end{equation}
and so one must solve 
\begin{equation}
    {d^2 \phi \over d\tau^2} + ({\bf \nabla}\phi)^2 
         = {dV\over d\phi} \, .
\label{flatscalar}
\end{equation}
The boundary conditions are that the path must start at the
homogeneous false vacuum configuration, with $\phi({\bf x}, \tau_{\rm
init}) = \phi_{\rm fv}$, and that $d\phi/d\tau = 0$ at $\tau_{\rm
fin}$ for all $\bf x$.  (Because $\phi_{\rm fv}$ is a minimum of the
potential, it turns out that $\tau_{\rm init} =- \infty$.)
A three-dimensional slice through the solution
at $\tau_{\rm fin}$ gives the most likely field configuration for the
nucleated true vacuum bubble.  This configuration, $\phi({\bf x},
\tau_{\rm fin})$, gives the initial condition for the subsequent
real-time evolution of the bubble.  As with the single-particle case,
a $\tau$-reflected solution is conventionally added to give a full
bounce.

Despite the fact that the spatial coordinates $\bf x$ and the path
parameter $\tau$ have very different physical meanings, there is a
remarkable mathematical symmetry in how they enter.  This suggests
looking for solutions that have an SO(4) symmetry; i.e., solutions for 
which $\phi$ is a function of only $s = \sqrt{{\bf x}^2 + \tau^2}$.
For such solutions, the field equation reduces to 
\begin{equation}
     {d^2\phi\over ds^2} + {3 \over s} {d\phi\over ds} 
        = {dV \over d\phi} \, .
\end{equation} 
The boundary conditions are 
\begin{equation}
   \left. {d\phi \over ds}\right|_{s =0} = 0 \,\,\,
      ,\qquad \phi(\infty)  = \phi_{\rm fv}   \, ,
\label{flatboundary}
\end{equation}
where the first follows from the requirement that the solution be
nonsingular at the origin, and the second ensures both that a
spatial slice at $\tau = -\infty$ corresponds to the initial state,
and that the slices at finite $\tau$ have finite energy relative to
the initial state.  Note that while $\phi(0)$ is required to be on the
true vacuum side of the barrier, it is not equal to (although it may
be close to) $\phi_{\rm tv}$.

Although the tunneling exponent is readily obtained from the WKB
approach, the prefactor, including (in principle) higher order
corrections, is most easily calculated from a path integral
approach~\cite{ColemanII}.  The
basic idea is to view the false vacuum as a metastable state with a
complex energy, with the imaginary part of the energy density yielding
the decay rate per unit volume.  The false vacuum energy is obtained
by noting that for large ${\cal T}$
\begin{equation}
    I({\cal T}) = 
    \int [d\phi] \, e^{-S_E(\phi)} \sim e^{-E_{\rm fv} {\cal T} } \, ,
\end{equation}
where the path integral is over configurations with 
$\phi({\bf x}, \tau= \pm {\cal T}/2) = \phi_{\rm fv}$.

This path integral can be calculated by summing the contributions from
the various stationary points, each of which gives a factor of $( \det
\,S'')^{-1/2}\, e^{-S}$.  Here $S''$ is the functional second
derivative of the action, evaluated at the stationary point; i.e., the
product of the frequencies of the normal modes.  The first stationary
point, a homogeneous false vacuum configuration with $\phi({\bf x},
\tau) = \phi_{\rm fv}$ everywhere, gives a contribution $Ae^{-S_{\rm
fv}}$, where the real prefactor $A$ includes the (properly
renormalized) determinant factor and $S_{\rm fv} = V(\phi_{\rm fv}) \,
{\cal T}{\cal V}$.  Here ${\cal V}$ denotes the volume of space and is
understood to be taken to infinity at the end of the calculation.

The next stationary point is the bounce solution to
Eq.~(\ref{flatscalar}).  The calculation of the determinant factor
here is complicated by the fact that $S''(\phi_{\rm bounce})$ has one
negative and four zero eigenvalues.  The former implies a factor of
$i$, which I will display explicitly.  The latter require the
introduction of collective coordinates; integrating over these gives a
factor of ${\cal T}{\cal V}$, corresponding to the fact that the
bounce can be centered anywhere in the four-dimensional Euclidean
space.

Finally, the approximate stationary points corresponding to
multibounce solutions also contribute, with the $n$-bounce contribution
including a factor of $({\cal T}{\cal V})^n/n!$ from
integrating over the positions of $n$ identical bounces.
Putting all this together gives a result that can be written as
\begin{eqnarray}
  I({\cal T})
  \!\! \!\! \!\! &=&\!\!\!\! \!\! A e^{- S_{\rm fv}} + i {\cal V}{\cal T} J
   e^{- S_{\rm bounce}} 
       + \cdots  \cr \cr 
   \!\!\!\! \!\! &=&\!\!\!\! \!\! A e^{- S_{\rm fv}} \left[ 1 + i {\cal
   V}{\cal T} J e^{-B}  
       + {1\over 2}\left(i {\cal V}{\cal T} J e^{-B} \right)^2 
      + \cdots \right]  \cr \cr
   \!\! \!\! \!\! &=&\!\!\! \!\!\! A e^{- {\cal T}{\cal V} V(\phi_{\rm fv})}
        \exp\left[i {\cal V}{\cal T} J e^{-B} \right]   \, .
\label{pathinteg}
\end{eqnarray}
Here $J$ includes both determinant and Jacobean factors, with the
latter arising from the introduction of the collective coordinates;
for present purposes, the important point is that it is real.
Extracting the energy density from the exponent in
Eq.~(\ref{pathinteg}) gives an imaginary part that is proportional to
${\cal V}$, corresponding to the fact that a bubble can nucleate
anywhere.  The quantity we actually want is the nucleation rate per
unit volume,
\begin{equation}
    \Gamma  = -{2 {\rm Im}\,E_{\rm fv} \over {\cal V}} 
         =  2J e^{-B}  \, .
\end{equation}

The path integral approach provides the vehicle for
extending~\cite{Langer:1969bc,Linde:1981zj} the
calculation to finite temperature $T$, with the path integral over
configurations extending from $ \tau = -\infty$ to $\tau = \infty$
replaced by one over over configurations that are periodic in $\tau$
with periodicity $1/T$.  At low temperature, where $1/T$ is larger
than the characteristic radius of the four-dimensional bounce, there
is little change from the zero-temperature nucleation rate.  However,
in the high-temperature regime where
$1/T$ is much smaller than this
characteristic radius the path integral is dominated by
configurations that are constant in $\tau$.  A spatial slice
at fixed $\tau$ gives a configuration, with total energy $E_{\rm
crit}$, that contains a single critical bubble. The exponent in the
nucleation rate takes the thermal form
\begin{equation}
    B = {E_{\rm crit}\over T} - {E_{\rm fv} \over T} \, .
\end{equation}
Note that, in contrast to the zero temperature case, there is no
spatial slice corresponding to the initial state.  Only through the
boundary conditions at spatial infinity does the bounce solution give
an indication of the initial conditions.

\section{Adding gravity}

Coleman and De Luccia~\cite{ColemanDeLuccia} argued that the effects
of gravity on vacuum decay could be obtained by adding an
Einstein-Hilbert term to the Euclidean action and then seeking bounce
solutions of the resulting field equations; as before, the tunneling
exponent would be obtained from the difference between the actions of
the bounce and the homogeneous initial state.  Their treatment did not
include the calculation of the prefactor, an issue that remains poorly
understood.

If one assumes O(4) symmetry, as in the flat spacetime case, the
metric can be written as 
\begin{equation}
     ds^2 = d\xi^2 + \rho(\xi)^2 d\Omega_3^2 \, ,
\end{equation}
where $d\Omega_3^2$ is the metric on the three-sphere, and the scalar
field depends only on $\xi$.  The Euclidean action becomes
\begin{eqnarray}
    S_E \!\! \!\! \!\!&=& \!\! \!\! \!\! 2\pi^2 \int d\xi \left[\rho^3 
        \left({1\over 2} \dot\phi^2 + V \right)
   \right.  \cr \!\! \!\! \!\!&& \!\! \!\! \!\!\qquad \qquad \left.
   + {3M_{\rm Pl}^2 \over 8\pi} 
        \left( \rho^2 \,\ddot \rho 
         + \rho\,\dot\rho^2 - \rho \right) \right] \, ,
 \end{eqnarray}
with dots denoting derivatives with respect to $\xi$.   The 
Euclidean field equations are
\begin{equation}
   \ddot \phi + {3 \dot\rho \over \rho} \, \dot\phi = {dV\over d\phi}
\label{curvedPhiEq}
\end{equation}
and
\begin{equation}
    {\dot \rho}^2 = 1 + {8 \pi \over 3 M_{Pl}^2}\,\rho^2
          \,\left({1\over 2} \dot\phi^2 - V \right) \, .
\label{rhoEq}
\end{equation}

\begin{figure}[th]
  \includegraphics[height=.28\textheight]{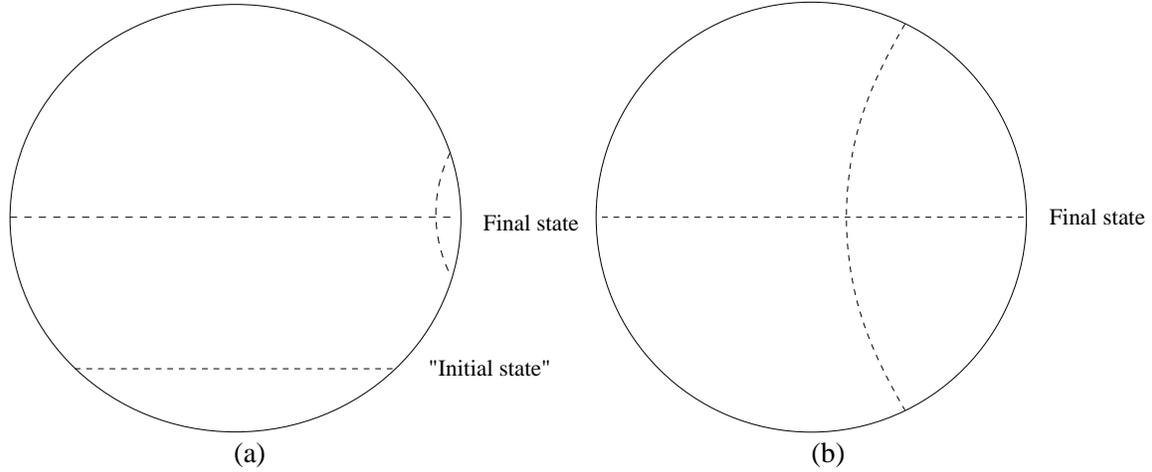}
  \caption{Schematic illustration of a Coleman-De Luccia bounce
 solution in two limiting regimes.  In both, $\phi$ is near its true
 vacuum value in the region to the right of the dashed arc, while on
 the left side it is near the false vacuum.  In both cases the
 equatorial slice denotes a three-sphere corresponding to the spatial
 hypersurface on which the bubble nucleates.  The lower dashed line in
 (a) represents a three-sphere indicative of the initial false vacuum
 state; this has no analogue in the regime illustrated in (b).}
  \label{bigsmallbounce}
\end{figure}

One can show that if $V(\phi)$ is everywhere positive, as I am
assuming here, then $\rho(\xi)$ has two zeros and the Euclidean space
is topologically a four-sphere.  One of the zeros of $\rho$ can be
chosen to lie at $\xi=0$, while the other is located at some value
$\xi_{\rm max}$.  Requiring the scalar field to be nonsingular then
imposes the boundary conditions
\begin{equation}
    \left.{d\phi \over d\xi}\right|_{\xi=0} = 0 \,, \,\,
  \qquad \left.{d\phi \over d\xi}\right|_{\xi_{\rm max}} = 0  \, .
\end{equation}
The symmetry of these boundary conditions should be contrasted with
the flat space boundary conditions of Eq.~(\ref{flatboundary}).  Note
that there is no requirement that scalar field ever achieve either of
its vacuum values, although $|\phi(\xi_{\rm max}) -\phi_{\rm fv}|$
is typically exponentially small
in cases where gravitational effects are
small.

Somewhat surprisingly, the Euclidean solution corresponding to a 
homogeneous false vacuum is not an infinite space, but rather a
four-sphere of radius 
\begin{equation}
    H_{\rm fv}^{-1} 
    = \sqrt{3M_{\rm Pl}^2\over 8\pi V(\phi_{\rm fv})}  \, .
\end{equation} 
Its Euclidean action is 
\begin{equation}
    S_{\rm E} = -{3 \over 8} {M_{\rm Pl}^4 \over
    V(\phi_{\rm fv})}  \, .
\end{equation}

If the parameters of the theory are such that the characteristic
radius of the 
flat space bounce
is much less than $H_{\rm fv}^{-1}$, then the curved space
bounce will be roughly as illustrated in Fig.~\ref{bigsmallbounce}a,
with the small region near $\xi =0$ corresponding to the true vacuum
region of the flat space bounce, the equatorial slice giving the
optimal configuration for emerging from the potential barrier, and a
slice such as that indicated by the lower dotted line roughly
corresponding to the state of the system before the tunneling process.
A bounce solution such as this yields a nucleation rate that only differs
only slightly from the flat space result.

On the other hand, there are 
choices of parameters that give a bounce
solution similar to that indicated in Fig.~\ref{bigsmallbounce}b, with
a true vacuum region that occupies a significant fraction of the
Euclidean space.  In this case, there is no slice that even roughly
approximates the initial state, suggesting that one should view this
as more analogous to a thermal transition in flat space than to 
zero-temperature quantum mechanical tunneling.

Indeed, for a bounce such as this the true and false vacuum regions
can perhaps be viewed as being on a similar footing, so that the
bounce can describe either the nucleation of a true vacuum bubble in a
region of false vacuum, or the nucleation of a false vacuum bubble in
a true vacuum region~\cite{Lee:1987qc}.  The rate for the former case
would be
\begin{equation}
   \Gamma_{{\rm fv}\rightarrow {\rm tv}}  \sim
   \exp\left\{-[S_{\rm E}({\rm bounce}) - S_{\rm E}({\rm fv})]\right\}
   \, ,
\end{equation}
while for the latter,
\begin{equation}
    \Gamma_{{\rm tv}\rightarrow {\rm fv}}  \sim 
   \exp\left\{-[S_{\rm E}({\rm bounce}) - S_{\rm E}({\rm tv})]\right\}
   \, .
\end{equation}
The ratio of these is
\begin{eqnarray}
    { \Gamma_{{\rm tv}\rightarrow {\rm fv}} 
        \over \Gamma_{{\rm fv}\rightarrow {\rm tv}} }
     \!\! \!\! \!\!&=& \!\! \!\! \!\!  \exp\left\{S_{\rm E}({\rm tv})
        - S_{\rm E}({\rm 
        fv})\right\} 
       \cr 
    \!\! \!\! \!\! &=& \!\! \!\! \!\! \exp\left\{{3 \over 8} {M_{\rm
        Pl}^4\over V(\phi_{\rm fv})} 
                -{3 \over 8} {M_{\rm Pl}^4\over V(\phi_{\rm tv}) }
        \right\}  \, .
\end{eqnarray}
If $V(\phi_{\rm fv}) - V(\phi_{\rm tv}) \ll V(\phi_{\rm fv})$, the
geometry of space is roughly the same in the two vacua, and we can
sensibly ask about the relative volumes of space occupied by the false
and true vacua.  In the steady state, this will be
\begin{eqnarray}
    {{\cal V}_{\rm fv} \over {\cal V}_{\rm tv}} 
        \!\! \!\! \!\!&=& \!\! \!\! \!\!
             { \Gamma_{{\rm tv}\rightarrow {\rm fv}} 
        \over \Gamma_{{\rm fv}\rightarrow {\rm tv}} } \cr
          \!\! \!\! \!\! &\approx&   \!\! \!\! \!\!
    \exp\left\{- {{4\pi \over 3} H^{-3}
      [V(\phi_{\rm fv}) -V(\phi_{\rm tv})] / T_{\rm dS}}\right\} \, .
\end{eqnarray}
The last line of this equation, which gives the ratio as the
exponential of an energy difference divided by the de Sitter
temperature, is quite suggestive of a thermal interpretation of
tunneling in this regime.

It is not hard to show that the flat space Euclidean field equations
always have a bounce solution.  This is no longer true when gravity is
included, as we will see more explicitly below.  However,
Eqs.~(\ref{curvedPhiEq}) and (\ref{rhoEq}) always have a homogeneous
Hawking-Moss~\cite{Hawking:1981fz} solution that is that is
qualitatively quite different from the flat space bounce.  Here $\phi$
is identically equal to its value $\phi_{\rm top}$ at the top of the
barrier, while Euclidean space is a four-sphere of radius $H^{-1}_{\rm
top} \equiv \sqrt{3M_{\rm Pl}^2/8\pi V(\phi_{\rm top})}$.  From this solution
one infers a nucleation rate
\begin{equation}
    \Gamma_{\rm fv} \sim
       \exp\left\{-{3 \over 8} {M_{\rm Pl}^4\over V(\phi_{\rm fv})}
         +{3 \over 8} {M_{\rm Pl}^4\over V(\phi_{\rm top})} \right\} \, .
\end{equation}
from the false vacuum

\begin{figure}[bt]
  \includegraphics[height=.56\textheight]{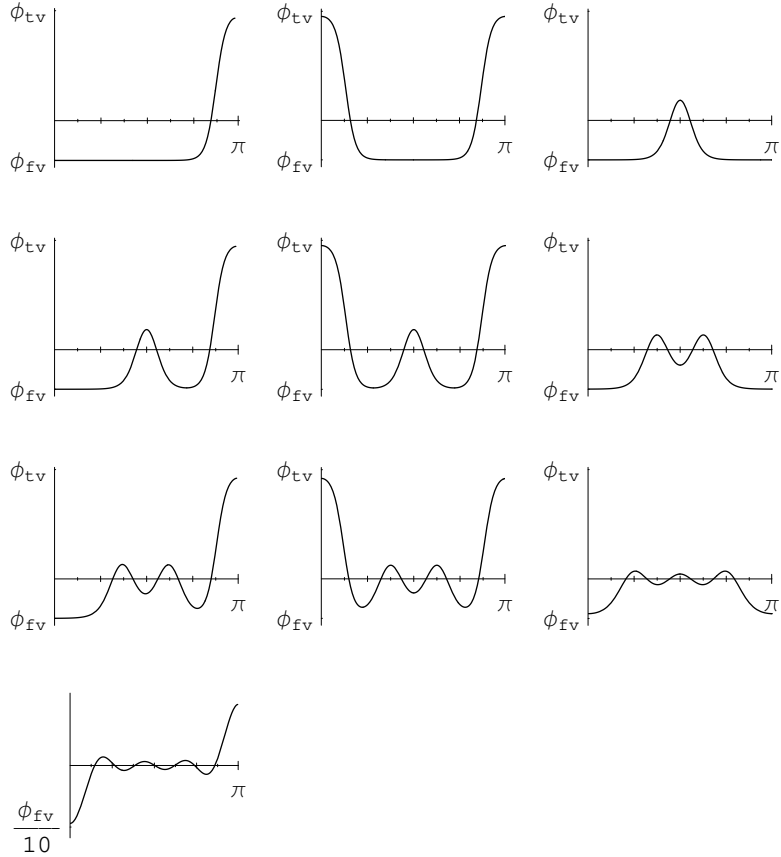}
  \caption{Bounce solutions for a scalar potential~\cite{Hackworth} 
   with cubic and quartic interactions and $\beta = 70.03$. }
  \label{array}
\end{figure}

\section{Other types of bounces?}

Given the existence of the Hawking-Moss solution, it is natural to
inquire whether the inclusion of gravity allows any other new classes
of Euclidean solutions.  In particular, might there be ``oscillating
bounce'' solutions
in which $\phi$ crosses the potential barrier not
once, but rather $k>1$ times, between $\xi=0$ and $\xi=\xi_{\rm max}$?

There can indeed be such solutions~\cite{Banks:2002nm}.
In examining their properties, we focussed on the case where $V(\phi_{\rm
top}) - V(\phi_{\rm tv}) \ll V(\phi_{\rm tv})$.  This simplifies the
calculations considerably, but does not seem to be essential for our
final conclusions.  With this assumption, the metric is, to a first
approximation, that of a four-sphere of fixed radius $H^{-1}$, and
$\xi_{\rm max} = \pi/H$.  We then only need to solve the scalar field
Eq.~(\ref{curvedPhiEq}).  Defining $y = H\xi$, we can write this as
\begin{equation}
    {d^2 \phi \over dy^2} + 3 \cot y {d\phi \over dy} = {1 \over H^2}
         {dV \over d\phi}  \, .
\end{equation}

It is convenient to start by first examining ``small amplitude''
solutions in which $\phi(0)$ and $\phi(\pi)$ are both close to
$\phi_{\rm top}$.  Let us assume that near the top of the barrier $V$
can be expanded as\footnote{The omission of cubic terms here is only
to simplify the algebra.  There is no difficulty, and little
qualitative change, in including such terms.  The details are given in
Ref.~\cite{Hackworth}.}
\begin{equation}
  \tilde V(\phi) =  V(\phi_{\rm top}) 
     - {H^2\beta \over 2} (\phi- \phi_{\rm top})^2 
      + {H^2\lambda \over 4} (\phi- \phi_{\rm top})^4
      + \cdots 
\end{equation}
with
\begin{equation}
    \beta = {| V''(\phi_{\rm top})| \over H^2}  \, .
\end{equation}
Keeping only terms linear in $(\phi-\phi_{\rm top})$ in
Eq.~(\ref{curvedPhiEq}) gives
\begin{equation}
   0 =  {d^2 \phi \over dy^2} + 3\cot y\, {d \phi \over dy} 
     + \beta (\phi - \phi_{\rm top}) \, ,
\end{equation}
whose general solution is
\begin{equation}
    \phi(y)- \phi_{\rm top} = A C_\alpha^{3/2}(\cos y) 
                 + BD_\alpha^{3/2}(\cos y)  \, ,
\end{equation}
where $C_\alpha^{3/2}$ and $D_\alpha^{3/2}$ are Gegenbauer functions
of the first and second kind and $\alpha(\alpha+3) = \beta$.  The
vanishing of $d\phi/dy$ at $y=0$ implies that $B=0$; the analogous
condition at $y=\pi$ is satisfied only if $\alpha$ is an integer, 
in which case $C_\alpha^{3/2}$ is a polynomial. 

While the linearized equation only has solutions for special values of
$\beta$, this condition is relaxed when the nonlinear terms are
included.  Furthermore, the nonlinear terms fix the amplitude of the
oscillations, which is completely undetermined at the linear level.
The problem can be analyzed by an approach similar to that used 
to treat the anharmonic oscillator.
Any function with $d\phi/dy$ vanishing at both $y=0$ and $y=\pi$
can be expanded as 
\begin{equation}
   \phi(y) = \phi_{\rm top} + {1 \over \sqrt{|\lambda|}} 
                \sum_{M=0}^\infty A_M C_M^{3/2}(y) \, .
\end{equation}
Substituting this into Eq.~(\ref{curvedPhiEq}) and keeping terms up to
cubic order in $(\phi-\phi_{\rm top})$ gives 
\begin{eqnarray}
   0 
   \!\! \!\! \!\!&=& \!\! \!\! \!\!
  \sum_{M=0}^\infty C_M^{3/2}(\cos y) \Big[ [\beta - M(M+3)] A_M 
     \cr 
   \!\! \!\! \!\! && \!\! \!\! \!\!\qquad
- {\rm sgn}\,(\lambda) \sum_{I,J,K} A_I A_J A_K q_{IJK;M}
    \Big]  \, ,
\label{phiExpansion}
\end{eqnarray}
where the $q_{IJK;M}$ arise from expanding products of three Gegenbauer
polynomials.
Requiring that the quantities multiplying each of the $C_M^{3/2}$ separately
vanish yields an infinite
set of coupled equations.  These simplify, however, if $|\Delta|
\equiv |\beta -N(N+3)| \ll 1$ for some $N$.  In this case, one
coefficient,  
$A_N$, is much greater than all the others.  The $M=N$ term
in Eq.~(\ref{phiExpansion}) then gives (to leading order)
\begin{equation}
  A_N = \pm \sqrt{ \Delta \over {\rm sgn}\,(\lambda) q_{NNN;N} } \, ,
\label{amplitude}
\end{equation}
where $q_{NNN;N} >0$.

If $\lambda>0$, Eq.~(\ref{amplitude}) only gives
a real value of $A_N$ if $\beta >
N(N+3)$.  As $\beta$ is increased through this critical value, two 
solutions appear.  These are essentially small oscillations about
the Hawking-Moss solution, with $\phi(0) \approx \phi_{\rm top} \pm
A_N C_N^{3/2}(1)$ and $\phi(\pi) \approx \phi_{\rm top} \pm A_N
C_N^{3/2}(-1)$.  Between these endpoints, $\phi$ crosses the top of
the barrier $N$ times.  If $N$ is even, the two solutions are
physically distinct, with one having $\phi$ on the true vacuum side of
the barrier at both endpoints, and the other having both endpoint
values on the false vacuum side.  If $N$ is odd, the two solutions are
just ``$y$-reversed'' images of each other.

As $\beta$ is increased further, the endpoints move down the sides of
the barrier, until eventually the small amplitude approximation breaks
down.  Nevertheless, we would expect the solutions to persist, with
$\phi(0)$ and $\phi(\pi)$ each moving toward one of the vacua.  When
$\beta$ reaches the next critical value, $(N+1)(N+4)$, two new
solutions, with $N+1$ oscillations about $\phi_{\rm top}$, will
appear, but the previous ones will remain.  Thus, for $N(N+1) < \beta
< (N+1)(N+4)$, we should expect to find solutions with $k = 0, 1, 2,
\dots, N$ oscillations.  We have confirmed these expectations by
numerically integrating the bounce equations for various values of the
parameters; the solutions for a typical potential with $\beta = 70.03$
are shown in Fig.~\ref{array}.

The fact that the number of solutions should increase with $\beta$ is
physically quite reasonable.  One would expect the minimum distance
needed for an oscillation about $\phi_{\rm top}$, like the thickness
of the bubble wall itself, to be roughly $|V''|^{-1/2}$.  Hence, the
number of
oscillations that can fit on a sphere of radius $H^{-1}$ should be of
order $H^{-1}/|V''|^{-1/2} = \sqrt{\beta}$.  In particular, this
suggests that for $\beta < 4$ there should not even be a $k=1$
Coleman-De Luccia bounce~\cite{Jensen:1983ac}.

It is thus somewhat puzzling to note the implications of
Eq.~(\ref{amplitude}) for the case where $\lambda$ is negative.  Here,
increasing $\beta$ through a critical value causes two solutions to
merge into the Hawking-Moss solution and disappear, suggesting that
the number of solutions is a decreasing function of $\beta$.  The
resolution to this can be found by analytically and
numerically examining various potentials that are unusually flat at the top.
In all the cases we have examined, the number of solutions is governed
by a parameter $\gamma$ that measures an averaged value of $|V''|/H^2$
over the width of the potential barrier.  When $\gamma$ is
sufficiently small, there are no bounce solutions (other than the
Hawking-Moss, which is always present).  
As $\gamma$ is
increased, new solutions appear at critical values.  These first
appear as solutions with finite values of $\phi(0) - \phi_{\rm top}$.
They then bifurcate, with $\phi(0)$ for one solution moving toward a
vacuum and $\phi(0)$ for the other moving toward $\phi_{\rm
top}$, eventually reaching it and disappearing when $\beta$ is at a
critical value.  The net effect is that the number of solutions
generally increases with $\gamma$, although it is not strictly
monotonic.

\section{Interpreting the oscillating bounces}

How should these oscillating bounce solutions be interpreted?  For the
flat space bubble, a spacelike slice through the center of the bounce
gives the initial conditions for the real-time evolution of the system
after nucleation; these predict a bubble wall with a well-defined
trajectory and a speed that soon approaches the speed of light.  The
interpretation of the Coleman-De Luccia bounce is similar.  The main
new feature here is the fact that the spacelike slice is finite.
Formally, this corresponds to the fact that de Sitter spacetime is a
closed universe, even though we expect the bubble nucleation
process to proceed similarly in a spacetime that only approximates de
Sitter locally.

The Hawking-Moss solution can be interpreted as corresponding to a
thermal fluctuation of all of de Sitter space (or, more plausibly, of
an entire horizon volume) to the top of the potential barrier.
Strictly speaking, classical Lorentzian evolution would leave $\phi$
at the top of the barrier forever.  However, this is an unstable
configuration, and so would be expected to break up, in a stochastic
fashion, into regions that evolve toward one vacuum or the other.

The oscillating bounce solutions yield a hybrid of these two extremes.
The endcap regions near $\xi=0$ and $\xi=\xi_{\max}$ clearly evolve
into vacuum regions analogous to those from the Coleman-De Luccia
bounce, while the intermediate, ``oscillating'', region is like that
emerging from a Hawking-Moss mediated transition.  As with the
Hawking-Moss solution, the bounce carries no information about the
initial state, and there is not even any correlation between the vacua
in the endcaps and the initial vacuum state.  Thus, like Hawking-Moss,
it is reminiscent of finite temperature tunneling in the absence of
gravity, and provides evidence of the thermal nature of de Sitter
spacetime.

The relative importance of the various solutions depends on the values
of their Euclidean actions.  Although the details vary with the
particular form of the potential, the various regimes are
characterized by a parameter $\gamma$ measuring an averaged value of
$|V''|/H^2$.  If $\gamma \gg 1$, there is a Coleman-De Luccia bounce,
a Hawking-Moss solution, and many oscillating bounces.  However, the
Coleman-De Luccia bounce has a much smaller action than the others,
and so dominates.  This is a regime of quantum tunneling transitions
followed by deterministic classical evolution.  At the other extreme
is the case where $\gamma \lesssim 1$, where the Hawking-Moss is the
only solution to the bounce equations.  This is a regime of thermal
transitions followed by stochastic real-time evolution.  In between is
a transitional region, with thermal effects still important.  It is
here that the oscillating bounces are most likely to play a role.

\begin{theacknowledgments}

This work was supported in part by the U.S. Department of Energy.

\end{theacknowledgments}

\end{document}